\documentclass[]{aa}

\usepackage{graphicx}
\usepackage[varg]{txfonts}
\usepackage{natbib}
\bibpunct{(}{)}{;}{a}{}{,} 
\usepackage[a4paper,breaklinks]{hyperref}

\def\teff{$T\rm_{eff}$}
\def\kms{$\mathrm{km\, s^{-1}}$}

\newcommand{\mygi}{MyGIsFOS}

\newcommand{\glog}{\ensuremath{\log g}}
\newcommand{\logg}{\ensuremath{\log g}}

\newcommand{\draftflag}{false}

\newcommand{\beq}{\begin{equation}}
\newcommand{\eeq}{\end{equation}}

\idline{10}{10}

\begin{document}

\title{TOPoS: IV. Chemical abundances from high-resolution observations of
seven EMP stars\thanks{Based
on observations obtained at ESO Paranal Observatory,
Programmes 189.D-0165,090.D-0306, 093.D-0136, 096.D-0468}
}

\author{
P.~Bonifacio  \inst{1} \and
E.~Caffau     \inst{1} \and
M.~Spite      \inst{1} \and
F.~Spite      \inst{1} \and
L.~Sbordone   \inst{2} \and
L.~Monaco     \inst{3} \and
P.~Fran\c cois\inst{1,4} \and
B.~Plez       \inst{5} \and
P.~Molaro      \inst{6}  \and
A. J.~Gallagher \inst{7,1} \and
R.~Cayrel     \inst{1} \and
N.~Christlieb \inst{8} \and
R.\,S.~Klessen     \inst{9} \and
A.~Koch,      \inst{10,11} \and
H.-G.~Ludwig  \inst{8,1} \and
M.~Steffen     \inst{12,1} \and
S.~Zaggia     \inst{13} \and
C.~Abate      \inst{14}
}

\institute{ 
GEPI, Observatoire de Paris, PSL Research University, CNRS, Place Jules Janssen,
92190 Meudon, France
\and
European Southern Observatory, Casilla 19001, Santiago, Chile
\and
Departamento de Ciencias Fisicas, Universidad Andres Bello, Fernandez Concha
700, Las Condes, Santiago, Chile
\and
UPJV, Universit\'e de Picardie Jules Verne, 33 Rue St Leu, F-80080 Amiens
\and
Laboratoire Univers et Particules de Montpellier, LUPM, Universit\'e de Montpellier,
CNRS, 34095 Montpellier cedex 5, France
\and
Istituto Nazionale di Astrofisica,
Osservatorio Astronomico di Trieste, Via G.B. Tiepolo 11, 34143 Trieste,
Italy
\and
Max Planck Institute for Astronomy, K\"onigstuhl 17, 69117 Heidelberg, Germany
\and
Zentrum f\"ur Astronomie der Universit\"at Heidelberg, Landessternwarte,
K\"onigstuhl 12, 69117 Heidelberg, Germany
\and
Zentrum f\"ur Astronomie der Universit\"at Heidelberg,
Institut f\"ur Theoretische Astrophysik, Albert-Ueberle-Stra$\beta$e 2, 69120
Heidelberg, Germany
\and
Zentrum f\"ur Astronomie der Universit\"at Heidelberg,
Astronomisches Rechen-Institut,  M\"onchhofstra$\beta$e 1214, 69120 Heidelberg,
Germany
\and
Department of Physics, Lancaster University, LA1 4YB, Lancaster, UK
\and
Leibniz-Institut f\"ur Astrophysik Potsdam (AIP), An der Sternwarte 16, 14482
Potsdam, Germany
\and
Istituto Nazionale di Astrofisica,
Osservatorio Astronomico di Padova, Vicolo dell'Osservatorio 5, 35122 Padova,
Italy
\and
Argelander Institut f\"ur Astronomie, Auf dem H\"ugel 71, 53121, Bonn, Germany
}
\authorrunning{Bonifacio, Caffau, Spite et al. }
\titlerunning{TOPoS IV}
\offprints{P.~Bonifacio}
\date{Received ...; Accepted ...}

\abstract%
{Extremely metal-poor stars provide
us with indirect information on the first generations of massive stars. 
The  TOPoS survey has been designed to increase the census of these
stars and to provide a chemical inventory that is  as detailed as possible.
}
{Seven of the most iron-poor stars have been observed
with the UVES spectrograph at the ESO VLT Kueyen 8.2\,m telescope
to refine their chemical composition. 
}
{We  analysed the spectra based on 1D LTE model atmospheres, 
but also used 3D hydrodynamical simulations of stellar atmospheres. 
}
{We measured carbon in six of the seven stars: all are carbon-enhanced
and belong to the low-carbon band, defined in the TOPoS II paper. 
We  measured lithium (A(Li)=1.9) in the most iron-poor
star (SDSS\,J1035+0641, [Fe/H]$< -5.2$). We were also able to measure Li
in 
three stars at [Fe/H]$\sim -4.0$,  two of which lie on the Spite plateau.

We confirm that SDSS\,J1349+1407 is extremely rich in Mg, but not in Ca.

It is also very rich in Na.
Several of our stars are characterised by low $\alpha$-to-iron ratios.
}
{The lack of high-carbon band stars  at low metallicity can be understood
in terms of evolutionary timescales of binary systems.
The detection of Li in  SDSS\,J1035+0641 places a strong constraint
on theories that aim at solving the cosmological lithium problem. 
The Li abundance of the two warmer stars at [Fe/H]$\sim -4.0$
places them on the Spite plateau, while the third, cooler star, lies below.

We argue that this suggests that the temperature at which Li depletion
begins increases with decreasing [Fe/H]. 
SDSS\,J1349+1407 may belong to a class of Mg-rich EMP stars. 
We cannot assess if  there
is a scatter in $\alpha$-to-iron ratios among the  EMP stars or if there
are several
discrete populations. However, the existence of stars with low $\alpha$-to-iron
ratios
is supported by our observations.
}
\maketitle


\section{Introduction}

Extremely metal-poor (EMP) stars are the shining witnesses of a primordial
Universe
corresponding to a redshift  higher than 10. 
They belong to the second generation of stars, objects formed from the gas
enriched
only by the first generation of the stars, Pop\,III stars.
Direct observation at high redshift of the individual
massive Pop\,III stars is not feasible because they are too 
faint for the current and next generation of telescopes.
There is one possible exception.
In the case of Pop\,III stars formed with masses of a few  hundred M$_\odot$
\citep{hirano14},
they could have ended their life with a pair-instability supernovae (PISN)
event.
According to \citet{bromm14}, this kind of event could be visible for the
James Webb Space Telescope (JWST) planned to be launched in 2019.
However, even in this event it is unlikely that the low-resolution spectra
that could be obtained from JWST would be able to provide a detailed chemical
inventory of PISN ejecta.
At the moment, EMP stars are  the only way to deduce the characteristics
of the Universe at the time they formed.
The TOPoS project \citep{topos1} has been designed to increase the number
and
chemical inventory of EMP turn-off stars. The TOPoS targets were selected
from the stars observed spectroscopically in the SDSS/SEGUE/BOSS survey
with the following colour cuts: $0.18 \le (g-z)_0 \le 0.70$ and $(u-g)_0
> 0.70$.
This allows the identification of  stars likely to be close to the main sequence
turn-off of halo stars.

\section{Observations and data reduction}

Target stars were observed using the ESO Ultraviolet and Visual Echelle Spectrograph
\citep[UVES,][]{dekker00} mounted at the Kueyen 8.2m telescope, the second
unit telescope
of the Very Large Telescope (VLT, Cerro Paranal, Chile). Observations of
five stars were collected in service
mode in the course of Large Programme 189.D-0165 (PI E. Caffau)
between April 2013 and March 2016. We adopted setting   Dic\,1 390+580  and
 therefore observed simultaneously with the blue and red UVES arms, centred
at 390\,nm and 580\,nm, and covering the spectral ranges 326-454\,nm and
476-684\,nm, respectively. The data were binned $2\times 2$ (spectral $\times$
spatial direction),
and a  1.4$^{\prime\prime}$ slit width was adopted. This resulted in a
resolving power of R$\simeq$30000 or larger when the image quality was better
than
1.4$^{\prime\prime}$. A different number of 3005\,s frames were taken for
six of the
target stars, namely 1, 5, 11, 5, 3, and 6 for stars SDSS\,J0140+2344,
SDSS\,J1034+0701, SDSS\,J1035+0641, SDSS\,J1247--0341,
SDSS\,J1442--0015, and SDSS\,J1507+0051, respectively. The 
signal-to-noise ratios per pixel of individual  spectra  varied in the range
3-30 for the blue arm and  8-61 for
the red arm.

\begin{figure*}
\begin{center}
\resizebox{\hsize}{!}{\includegraphics[draft = \draftflag, clip=true]
{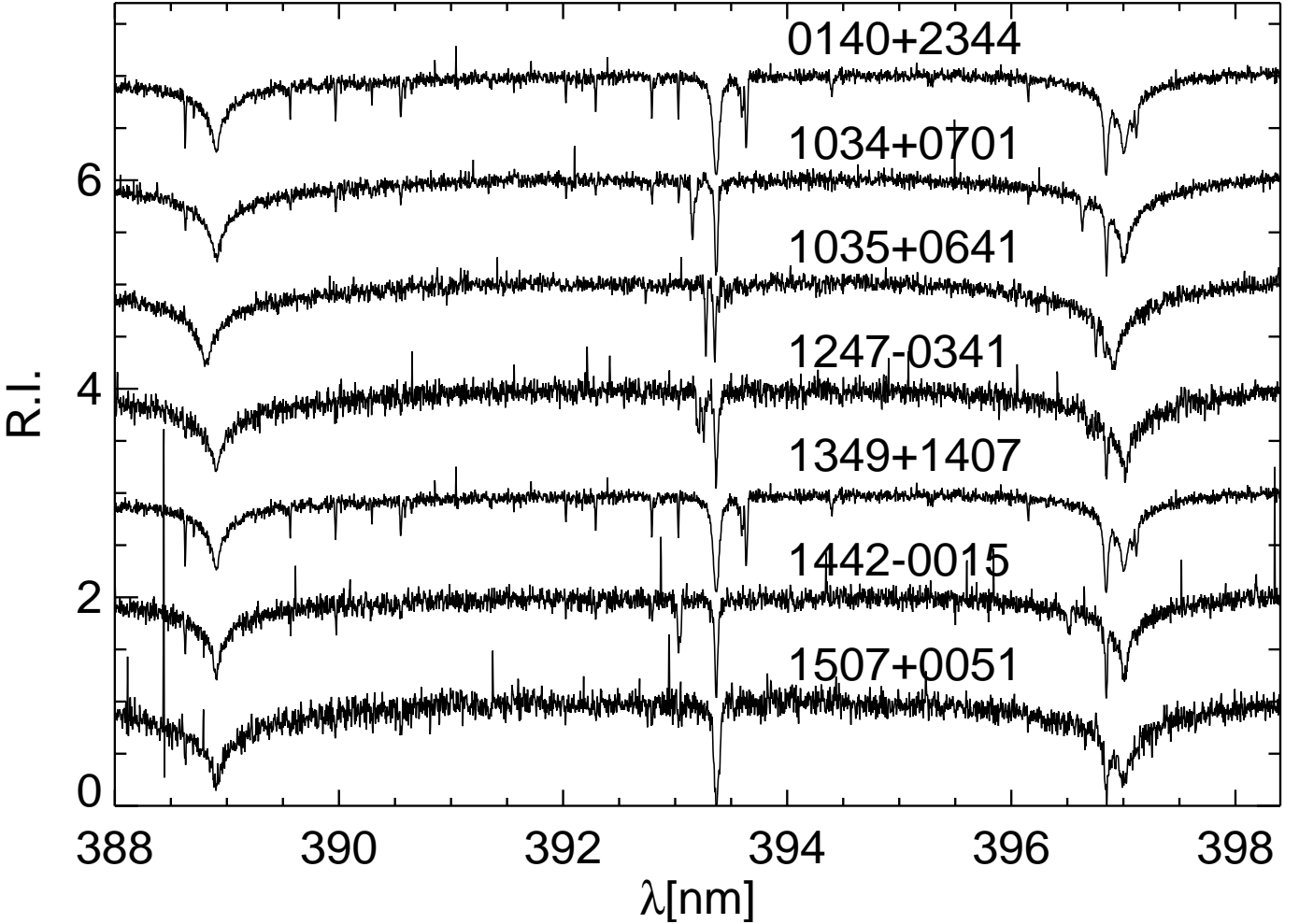}}
\end{center}
\caption[]{Spectral ranges of the \ion{Ca}{ii}-H and -K lines  of the sample
stars.
}
\label{plot_HK}
\end{figure*}

\begin{figure*}
\begin{center}
\resizebox{\hsize}{!}{\includegraphics[draft = \draftflag, clip=true]
{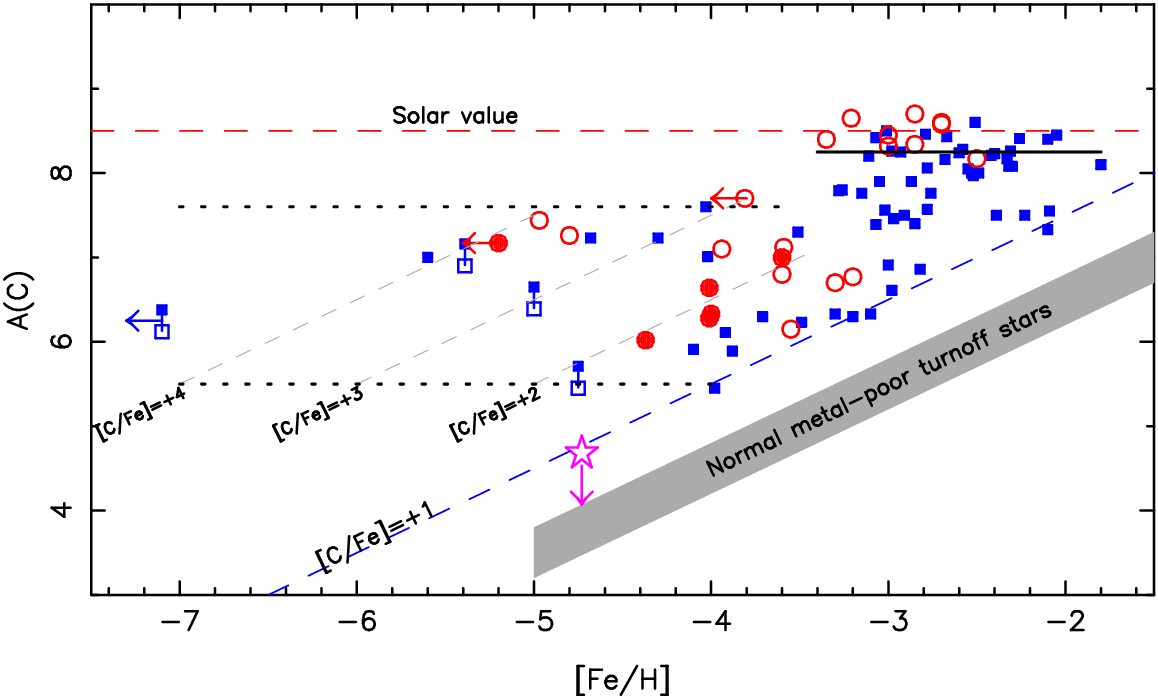}}
\end{center}
\caption[]{ Carbon abundances of the programme stars (filled red circles)

as a function of [Fe/H]. Other stars
published by our group are shown as open red circles. 
Other stars from the literature
\citep{sivarani,Plez05,PCM05,frebel05,frebel06,thompson08,aoki08,behara,masseron10,masseron12,yong2013,Cohen13,Li2015}
are shown as blue squares. 
The violet star is the upper limit on  SDSS\,J1029+1729 \citep{leostar,leostaraa}.
The horizontal dotted lines delimit the low-carbon band.
}
\label{plot_c}
\end{figure*}

\begin{table}
        \caption{ SDSS DR12 $^1$ photometry and exposure times for the programme
stars.}
\label{data}
\begin{center}
\renewcommand{\tabcolsep}{1pt}
\begin{tabular}{lccccccccr}
\hline \noalign{\smallskip}
 Star                   & $u$      & $g$      & $r$      & $i$     & $z$
     & T$_{exp}$ & N$_{\rm exp}$\\
                        &           &           &           &           &
          & s             & \\                                          
               
\hline \noalign{\smallskip}                                             
             
SDSS\,J014036.22+234458.0  & 16.78 & 15.82 & 15.35 & 15.12 & 15.03 & 3005
& 1 \\
SDSS\,J103402.71+070116.6  & 18.34 & 17.52 & 17.23 & 17.13 & 17.12 & 3005
& 5 \\
SDSS\,J103556.11+064143.9  & 19.53 & 18.65 & 18.37 & 18.31 & 18.26 & 3005
& 29 \\
SDSS\,J124719.46--034152.4 & 19.32 & 18.50 & 18.24 & 18.15 & 18.14 & 3005
& 5 \\
SDSS\,J134922.91+140736.9  & 17.44 & 16.65 & 16.33 & 16.24 & 16.19 & 3600
& 9  \\
SDSS\,J144256.37--001542.8 & 18.76 & 17.96 & 17.65 & 17.51 & 17.49 & 3005
& 3 \\
SDSS\,J150702.02+005152.6  & 19.75 & 18.77 & 18.57 & 18.48 & 18.41 & 3005
& 6 \\
\noalign{\smallskip}
\hline
\end{tabular}
$^1$In this work we used the SDSS DR12 photometry; the magnitudes of the

        latest release DR14, differ from these values by less that 0.015
mag.
\end{center}
\end{table}

Star SDSS\,J1035+0641 was also observed with setting Dic\,2 437+760
(programme ID: 096.D-0468 PI Sbordone), with the blue and red arms covering
the spectral ranges
373-499\,nm and 565-946\,nm, respectively. A total of 18\,$\times$3005\,sec
frames were acquired. The same slit width (1\farcs{4}) and on-chip
binning ($2\times 2$) were chosen. Signal-to-noise ratios per pixel of individual
spectra  varied in
the range 5-10 and 4-8 for the blue and red arms, respectively.

SDSS\,J1349+1407  was observed in service mode in the course
of programmes 090.D-0306 and 093.D-0136 (PI Sbordone) between March 2013
and
June 2014. The setting used in this case was  Dic2 437+760 with
a $1\times 1$ binning and a 1\farcs{0} slit, providing a resolving power
of R$\sim 48000$. A total of nine integrations
of 3600s each were acquired. The coadded spectrum attains a S/N $\sim 40$
per pixel at 400\,nm
and a S/N $\sim 60$ per pixel at 590\,nm. 

The data were reduced by the ESO staff, and retrieved through the Phase\,3
spectral
data products query
form\footnote{\url{http://archive.eso.org/wdb/wdb/adp/phase3_spectral/form}}.

\section{Analysis}

The chemical analysis has been done with the code \mygi\ \citep{mygisfos}.
This code derives the chemical abundances by fitting a selected set of features.
The best fit profile is obtained by interpolating in a series of pre-calculated
synthetic spectra.
These synthetic spectra have been computed with {\tt turbospectrum} \citep{alvarez_plez,2012ascl.soft05004P}
and are based on a grid of 1D plane-parallel hydrostatic model atmospheres
previously computed with MARCS \citep{G2008}.
The procedure to derive the  atmospheric parameters and chemical composition
is described by \citet{topos1} and has
also been  used in the other papers of this series \citep{topos2,topos3}.
There were  two exceptions for which \mygi\ was not well adapted:    carbon
in all stars, and magnesium
in the star SDSS\,J1349+1407. Carbon  was derived by $\chi$-squared fitting
of the 
G-band with a specifically computed grid of synthetic spectra with fixed
metallicity, \teff,\ and \glog\ 
and with varying C abundance. This technique is the same as that
used in our previous papers \citep{spite13,topos2}. 
To take into account the known 
granulation effects on the G-band \citep{Collet2007,behara,Gallagher2016,Gallagher2017},
we used the corrections provided by \citet{Gallagher2016}, assuming a metallicity
of --3.0.
The stellar parameters and abundances of C and Li are provided in Table\,\ref{param}.

Because
its [Mg/Fe] ratio is so extreme (Mg enhanced by more than 1\,dex over Fe),
one concern for  SDSS\,J1349+1407 was that its temperature 
structure could be different
from
that of a normal solar-scaled and $\alpha$-enhanced model.
Magnesium
is one of the
main electron donors for solar-type stars, and its abundance may affect the
ionisation
structure of the atmosphere. 
We verified that the high [Mg/H] does not affect the abundances derived with
models enhanced in Mg by 
the canonical 0.4\,dex
by computing both the ATLAS\,9 and ATLAS\,12 models \citep{kurucz93,kurucz05,SBC04,sbordone05},
the latter with [Mg/Fe]=+1.4 to 
derive
the abundances.
From this test we conclude that 
the differences are negligible;  the largest difference reaches 0.05\,dex
for Mn.
At the very low metallicities relevant for the stars
analysed in this paper, Mg, Ca, and any metals cease to be important electron
donors;
as already discussed by \citet{nc04},
the main electron donor is, in fact, hydrogen.

\subsection{Individual stars and comparison with previous  analyses}

All  of our targets have already been analysed by our group
on the basis of X-Shooter spectra, and some have also been analysed  by other
groups.

\begin{itemize}
\item SDSS\,J0140+2344.
The stellar parameters and [Fe/H] we derive are in excellent agreement with
those determined
by \citet{topos1} from the analysis of an X-Shooter spectrum.
Our estimated upper limit of A(C) from the X-Shooter spectrum was too optimistic:
the carbon abundance we derive in this work is 0.3\,dex larger than the X-Shooter
upper limit. 

The Mg and Si abundances
measured from the X-Shooter and UVES spectra are in excellent agreement.

The Ca abundance measurements are more problematic.
We detect two \ion{Ca}{i} lines, the resonance line at 422.67\,nm and the
subordinate line 
at 442.54\,nm. The two lines are very discrepant, as is generally the case
in EMP stars, even when NLTE effects are taken into 
account \citep[see][and references therein]{spite12}.
The  422.67\,nm resonance line provides $[{\rm Ca/H}]=-3.97$, in good agreement
with the X-Shooter
spectrum where  the only \ion{Ca}{i} line detected was in fact the 422.67\,nm
line, 
                which implied $[{\rm Ca/H}]=-3.78$.
The subordinate line yields a Ca abundance that is over 1\,dex higher 
($[{\rm Ca/H}])=-2.79$). In Table\,\ref{abbo2} we only provide  the value
from the subordinate line,
which, as discussed by \citet{spite12}, is more reliable. 
The \ion{Ca}{ii} is derived from the K-line alone and it is a factor of three
lower  than
the value we had estimated from the X-Shooter spectrum; however, in that
case we relied only on the \ion{Ca}{ii} infrared triplet lines. The  abundance

from the K-line is also  strongly discrepant from
the \ion{Ca}{i} subordinate line.    
\citet{yong2013} analysed a  Keck/HIRES spectrum of this star. They derived
\teff = ${5703}$\,K from a combination  of spectrophotometry and
Balmer line analysis. This is in agreement, within the errors, with our
adopted \teff. The surface gravity was derived from isochrones and two 
solutions were found, depending on whether the star is a main sequence
or a subgiant  star, this corresponds to \glog = ${3.36}$
or \glog = 4.68. In the two cases, they derive [Fe/H]=$-4.11$ or
[Fe/H]=$-4.02$ (with our adopted solar Fe abundance \citealt{abbosun}), 
in excellent agreement with our result.
For Ca, they only measure the \ion{Ca}{i} 422.67\,nm resonance line
and derive [Ca/H]=$-3.82$ or [Ca/H]=$-3.88$, again in excellent
agreement with our results for the same line.

\citet{aoki13} analysed  a Subaru spectrum of this star
and used  \teff = ${6103}$\,K, which  was derived from the SDSS low-resolution
spectrum by the SEGUE Stellar Parameter Pipeline \citep[SSPP,][]{SSPP},
even though both $V-K$ and $g-r$ imply a temperature close to
that used by us. For the surface gravity they assumed 
\glog = 4.0 and, like us, they derived [Fe/H]=$-3.67$ (on our adopted
solar abundance scale).  Considering their adopted higher effective
temperature, this metallicity is consistent  both with ours and with that
of \citet{yong2013}. For Ca they derive [Ca/H]=$-3.82$ from a single line.
Although it is not explicitly stated, it is highly probable that this
refers to the   \ion{Ca}{i} 422.67\,nm resonance line. 

Finally, \citet{aguado16} analysed the SDSS spectrum and a  ISIS WHT spectrum
and derived
the stellar parameters ${6090\pm 200}$\,K/${4.7\pm 0.3}$/$-3.6\pm 0.2$.
Unsurprisingly, the \teff\ is close to the SSPP value and the derived metallicity
is consistent.

We conclude that the different analyses of this star are all consistent,
but we point
out the discrepancy of the Ca abundance derived from different lines. This
will have
to be revisited in the light of 3D NLTE computations in the future.
 
\smallskip
\item SDSS\,J1034+0701.
The  X-Shoooter spectrum is analysed in \citet{pfr17}. 
The [Fe/H] from the UVES spectrum is about 0.4\, dex lower than the value
derived from the X-Shooter spectrum.
The Mg abundance is in  agreement within errors.
We did not derive the Ca abundance from the UVES spectrum, but the Ca abundance
from the X-Shooter spectrum, based on the \ion{Ca}{ii}  IR triplet, and the
X-Shooter based Fe abundance implies [Ca/Fe]=+0.33.

\smallskip
\item SDSS\,J1035+0641.
\citet{topos2} investigated an X-Shooter spectrum, but they could only derive
A(C) and an upper limit for the Fe abundance.
The A(C) derived from the UVES spectrum
is a factor of two higher than the value we derived from the X-Shooter spectrum,
which illustrates
the limit of measuring the G-band for a warm star like this at moderate resolution.
The non-detection of the 382.0425\,nm line  (equivalent width $< 0.4$ pm
at $1 \sigma$)
provides an upper limit of the Fe abundance [Fe/H]$< -5.2$, tighter than
the value 
we derived from the X-Shooter spectrum.
A fit to the \ion{Ca}{ii} K line
provides an abundance that is consistent with that derived from the X-Shooter
spectrum.  

We found a mistake in Table 7 of \citet{topos2} and in their Figure 10. 
The upper limit on lithium, corresponding to an upper limit of 1.8\,pm, 
is A(Li)$< 2.05$ and not 1.1, as incorrectly listed in Table 7. Our measured
equivalent
width of 1.3\,pm is consistent with that upper limit and corresponds to A(Li)=1.90.

\smallskip
\item  SDSS\,J1247--0341. 
The metallicity determined by \citet{topos1} from the X-Shooter spectrum
is in good agreement with that derived from the UVES spectrum.
Also in this case the C abundance derived from
the UVES spectrum is larger by 0.6\,dex than the value we estimated
from the X-Shooter spectrum. The Mg abundances from X-Shooter and UVES spectra
are consistent
within 1$\sigma$. \citet{topos1} provided the Ca abundance from 
two lines of the \ion{Ca}{ii} IR triplet, which we considered more reliable
than the \ion{Ca}{i} 422.67\,nm resonance line. That line implied [Ca/H]=$-3.96$,
which is consistent with the abundance provided here for the same line. 
Also in this case we point out how different lines provide different abundances
for Ca; this is likely due to the inadequacy of the 1D LTE analysis of this
element
in these extremely metal-poor stars. 
The Gaia data release 1 \citep{GaiaDR1} shows that this star
has a close-by companion at a distance of 0\farcs{01} and 0.5 mag
fainter in $G$. Certainly the light of this companion contaminates our
spectrum, but since we do not have any colour information at this time we
did not
attempt to correct our spectrum for the veiling.
We do not detect any sign of the spectrum of the companion star in our spectrum.

\smallskip
\item SDSS\,J1349+1407. \citet{sbordone12} announced it  as a Mg-rich star.
The quality of the spectrum available at that time was sufficient only for
giving abundances of a few elements.
The star was independently recovered as a Mg-rich star by 
\citet{li14}, who analysed the SDSS\,DR9 spectra.
Their temperature for the star is higher by 230\,K and consequently their
metallicity is higher ([Fe/H]=--2.83).
In the UVES spectrum of SDSS\,J1349+1407 
we identified six \ion{Mg}{i} lines in the blue spectrum (382.9, 383.2, 383.8,
405.7, 416.7, 470.2\,nm).
We fitted the line profiles and
we derived ${\rm A(Mg)}=5.30\pm 0.16$ and by removing the line  which is
in the wing of a Balmer line,
we derive ${\rm A(Mg)}=5.36\pm 0.11$.
The star is also enhanced in Na, with [Na/Fe]=+0.86.

\smallskip
\item SDSS\,J1442--0015. 
We compared our spectrum with the X-Shooter spectrum of \citet{gto12}.
The metallicities from the analysis of the two spectra
are in good agreement within errors of less than $1 \sigma $. The Mg abundances
are in reasonable agreement with $1.5 \sigma$ errors. 
The situation is slightly worse for the Ca abundance; the 
abundances derived from the \ion{Ca}{i}
422.67\,nm resonance line from the two spectra are consistent to within
$1.8 \sigma$, i.e. 0.7\,dex. As usual, this is discrepant with the \ion{Ca}{ii}
IR
triplet lines, measured in the X-Shooter spectrum.
This example is a recommendation  not to  overinterpret the abundances
that rely on a single, weak line.

It is important to note that both here and in the study of \citet{gto12}

we adopted the effective temperature derived by fitting the wings of H$\alpha$,
which 
is considerably lower than  the temperature implied by the $g-z$ colour (6161\,
K).

\smallskip
\item SDSS\,J1507+0051. The X-Shooter spectrum  was analysed by \citet{topos1}.
Both the metallicity and the Mg abundance of the  two analyses 
are in agreement within less than $1 \sigma$. Instead, we have a strong discrepancy

for the abundances of Ca derived from \ion{Ca}{ii} lines. 
In the UVES spectrum we detect the 370.6024\,nm line, while in the X-Shooter
spectrum, we relied on the IR triplet lines. This discrepancy needs to be
further investigated.
\end{itemize}

\begin{figure}
\begin{center}
\resizebox{\hsize}{!}{\includegraphics[draft = \draftflag, clip=true]
{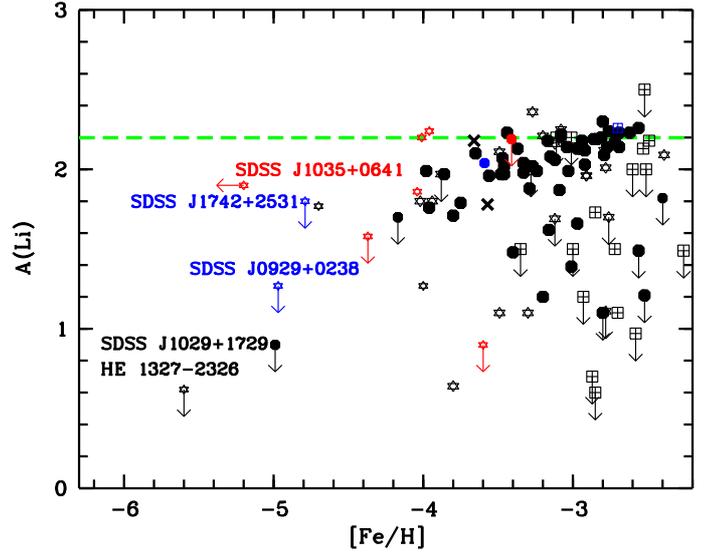}}
\end{center}
\caption[]{Lithium abundance in unevolved extremely metal-poor stars.
The different symbols refer to different carbon abundances. 
The filled hexagons refer to carbon normal stars.
CEMP stars of the low- and high-carbon bands are shown as star symbols and
crossed squares, respectively.
Measurements and upper limits of the programme stars are shown in red. 
Measurements and upper limits from our group's previous papers \citep{topos2,topos3}
are shown in blue.  
Black symbols  are stars for which metallicity, lithium abundance,
and carbon abundance are taken from the literature
\citep{norris97,Lucatello03,
sivarani04,Ivans05,sivarani,frebel07,frebel08,thompson08,aoki08,
sbordone10,behara,leostaraa,
Carollo12,masseron12,aoki13,
Ito13,Carollo13,spite13,Roederer14,Aoki15,sdss_uves,Li2015,terese,topos3,
Placco16,Matsuno17}.
The two components of the binary system CS 22876-32 \citep{jonay08} are shown
as black crosses.
The green dashed line is the level of the {\em} Spite plateau
as determined by \citet{sbordone10}.
\label{plot_Li}
}
\end{figure}

\begin{figure}
\begin{center}
\resizebox{\hsize}{!}{\includegraphics[draft = \draftflag, clip=true]
{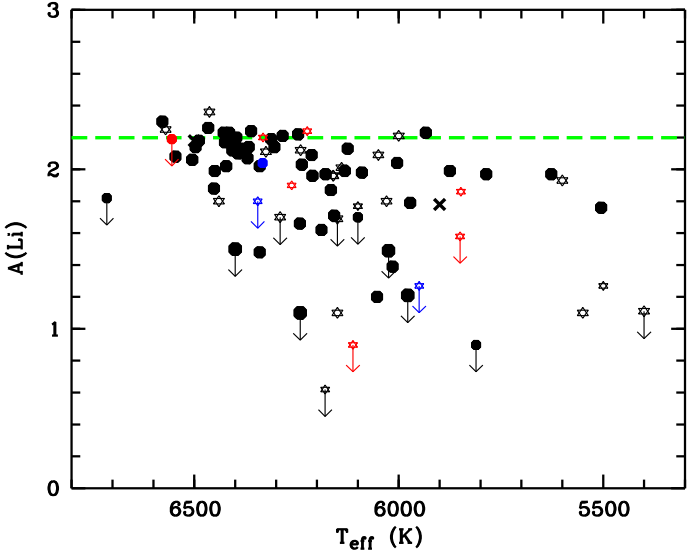}}
\end{center}
\caption[]{ Lithium abundance in the unevolved extremely metal-poor stars
as a function of effective temperature. The meaning of the symbols
is the same as in Fig.\,\ref{plot_Li}. The high-carbon band CEMP stars have
not been included in this plot.
\label{plot_Li_teff}
}
\end{figure}

\begin{figure}
\begin{center}
\resizebox{\hsize}{!}{\includegraphics[draft = \draftflag, clip=true]
{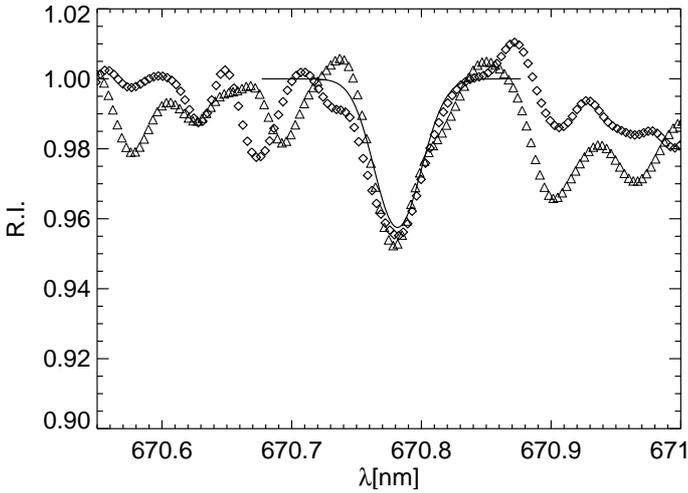}}
\end{center}
\caption[]{ Lithium doublet in SDSS\,J1034+0701 (squares) and
SDSS\,J1247--0341 (triangles) compared to a 3D NLTE theoretical profile
(\teff = 6270\,K \glog = 4.0 [M/H]=--3.0, A(Li)=2.1). The observed spectra
have been smoothed with a Gaussian of
4 \kms\ FWHM to increase the S/N. 
\label{plot_lim4}
}
\end{figure}

\section{Results and discussion}

The main result of this investigation is the confirmation, based on
higher resolution spectra, of the very low metallicities that we derived
for these stars from the analysis of the X-Shooter spectra. Two stars have
[Fe/H] below $-4.3$, 
three stars around $-4.0$, and two stars around $-3.5$. These numbers confirm
the high efficiency of the TOPoS strategy for target selection. 
The five stars with [Fe/H]$\le -4.0$ discussed in this paper, 
SDSS\,J1742+2531 ( [Fe/H]=$-4.80$, \citealt{topos2} ) and
SDSS\,J0929+0238 ([Fe/H]=$-4.97$, \citealt{topos3})
are the most iron-poor stars we found in the course of the TOPoS project
and they are all strongly C-enhanced. To date, among the stars with
[Fe/H]$\le -4.5$ the only `non C-enhanced star' found is  SDSS\,J1029+1729
\citep{leostar,leostaraa}.

\subsection{Carbon abundances}
It is interesting to note that all the C-enhanced stars that we have found
belong
indeed to the low-carbon band discussed by \citet{topos2}, as illustrated
in Fig.\,\ref{plot_c}.
These stars do not seem to be enhanced in $s$-process elements and we suggest
that they are indeed 
CEMP-no stars.  This view is supported also by the 
recent study of \citet{Hansen16} who analysed a sample of 27 metal-poor stars
and found that 20 of them
are CEMP, 
3 of which are CEMP-no
stars  that  belong to the low-carbon
band.
We  suggest here that a useful classification of metal-poor stars  
can be made using only their C abundance without any reference
to their abundance of n-capture elements. 
This is related to the fact that, for unevolved stars, it is very difficult
to secure data quality high enough
to derive measurements or significant upper limits for the heavy elements.
Our proposed classification scheme is as follows:
\begin{itemize}
\item `carbon normal': for [Fe/H]$\ge -4$ [C/Fe]$< 1.0$,
for  [Fe/H]$< -4$ A(C)$<5.5$;
\item low-carbon band CEMP stars: stars that do not fulfil the carbon normal
criterion and have A(C)$\le 7.6$;
\item high-carbon band CEMP stars: stars that do not fulfil the carbon normal
criterion and have  A(C)$ > 7.6$.
\end{itemize}

This classification is qualitatively similar to that proposed by \citet{Yoon},
except that their Group II is partly included in our
low-carbon band and mostly in our carbon normal  stars, their
Group I is by and large coincident to our high-carbon band, except for the
stars with the lowest C abundances in their Group I, which we assign to the
low-carbon
band.

In \citet{topos2} we also used the working hypothesis that all the stars
on the high-carbon band
are binaries and the high carbon abundance is the consequence of mass-transfer
from a companion
during its AGB phase. This hypothesis, however, does not place any lower
limit to the [Fe/H]
of the high-carbon band stars. The fact that, observationally, we do not
find
any high-carbon band stars below [Fe/H]=--3.6 is intriguing. In Fig.\,\ref{plot_c}
we have 26 stars with a metallicity below --3.6, of which 25 are C-enhanced,
and none is a high-carbon
band star. It therefore seems unlikely to admit that this is just the result
of small number statistics. 
The binary systems that can give rise to a high-carbon band star, in our
hypothesis, are severely constrained regarding the masses of the components.

In particular, the primary star should be larger than about $0.9 M_{\odot}$
to experience efficient third-dredge-up episodes during the AGB phase
\citep{stancliffe08, karakas10}, and consequently enrich its surface with
the carbon produced in the stellar interior. 
The secondary should be less massive than about $0.8-0.85 M_{\odot}$, because
higher-mass stars would become white dwarfs
on a timescale shorter than $10-12$ billion years, which is the approximate
age of the Galactic halo. Finally, mass transfer 
in binary systems with AGB donors is efficient in a limited range of orbital
periods between approximately $10^2$ and $10^5$ days \cite[e.g.][]{Izzard,
Abate}.

\citet{CL2008} computed the yields of AGB stars of extremely low  and zero
metallicity,
and they found that the [C/Fe] ratios of their yields increase
monotonically with decreasing [Fe/H] (as illustrated in their Figure 5).
The carbon abundance in the ejecta
of the AGB star varies within  a factor of four \relax  in the metallicity
range $-6.5\le$[Fe/H]$\le -3.0$ 
\citep{CL2008}.  Therefore, theoretically, there is no indication that
the nucleosynthetic yields of AGB stars  vary in metallicity in such 
a way as to disfavour the formation of high-carbon band stars.  
A recent review on the nucleosynthesis of AGB stars is given
by \citet{Karakas}.

\begin{figure*}
\begin{center}
\resizebox{\hsize}{!}{\includegraphics[draft = \draftflag, clip=true]
{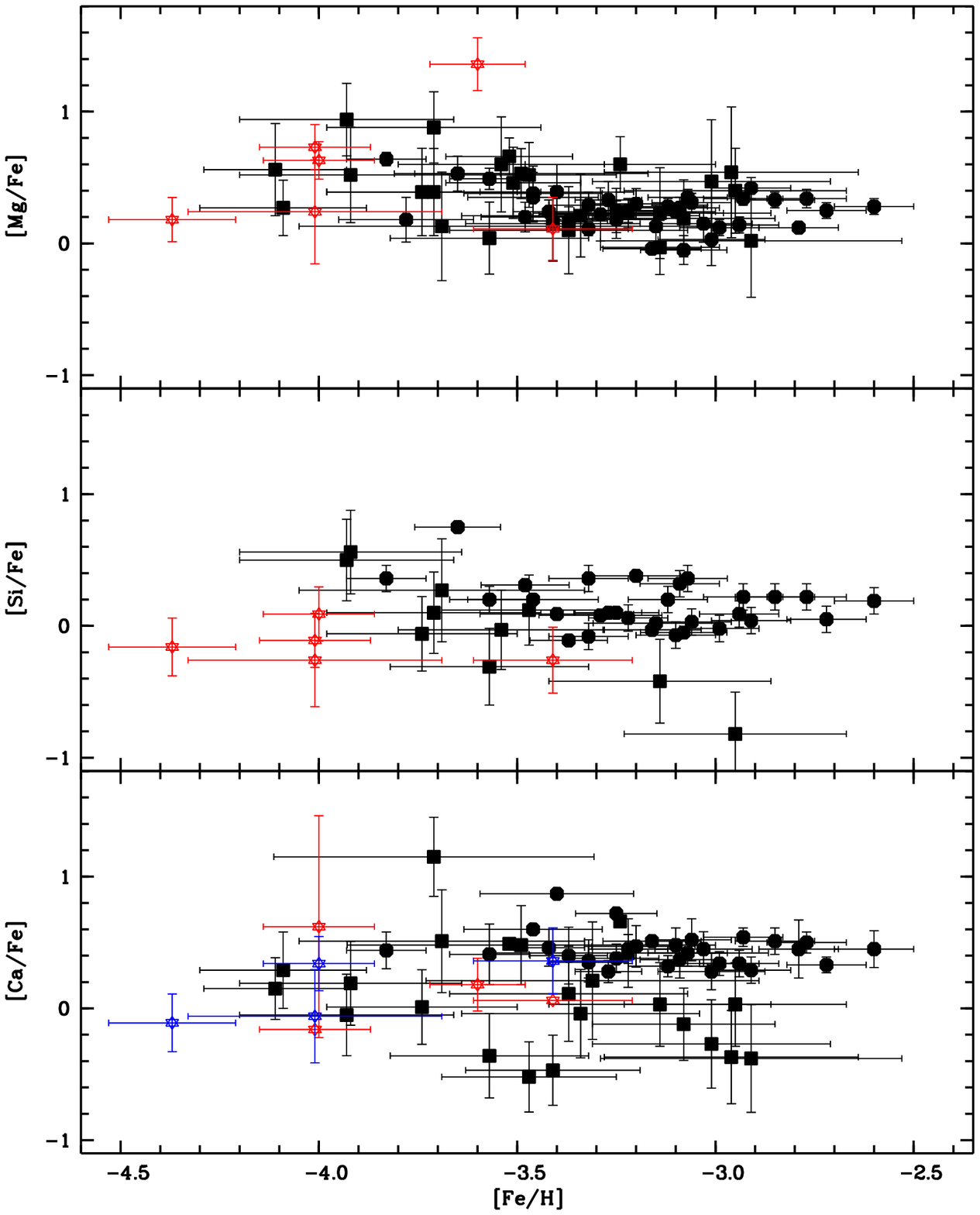}}
\end{center}
\caption[]{Ratio of $\alpha$ elements to iron
for the programme stars (red stars) compared to those measured
by our group in other extremely metal-poor dwarf stars \citep{bonifacio09,gto11,sdss_uves,gto12,topos1}.
Black squares are measurements from X-Shooter spectra, while black hexagons
are
measurements from UVES spectra. In the panel giving the [Ca/Fe] ratio, the
blue star symbols 
are \ion{Ca}{ii} measures. We restricted the x-axis so as to exclude
SDSS\,J1035+0641; which only has an upper limit on iron and thus a lower
limit on [Ca/Fe] and is not informative. 
}
\label{plot_alpha}
\end{figure*}

A possible explanation of the lack of high-carbon band stars at low [Fe/H]
is that at these
metallicities, such systems either do not form at all or are exceedingly
rare. 
The existence of a rather sharp cut-off  suggests the existence of a physical
mechanism that prevents the formation of these systems. 
Because the enriched material synthesised by the AGB star needs to be expelled
in the stellar wind 
and accreted by a low-mass companion, the lack of high-carbon band stars
could indicate that at low [Fe/H]
binary mass-transfer processes become rarer or less efficient. Although the
binary-star fraction is not 
well constrained at low metallicity, there is no evidence that it decreases
\cite[e.g.][]{gao14}. 
The data on the binarity among stars with
[Fe/H]$<-3.6$ is scanty, yet the discovery of the binarity
of SDSS\,J0929+0238 \citep{topos3} suggests that it cannot
be much lower than among stars of higher metallicity.
It has been suggested that at low metallicity the ejection of the AGB envelope
becomes increasingly 
inefficient \cite[][]{wood11}. If only a small mass is ejected by the primary,
the amount of 
material accreted by the secondary may be insufficient to enhance the carbon
abundance up to 
a typical high-carbon band value. 

Another effect to take into account is that when a star accretes
material and becomes more massive, it evolves more quickly than a single
star of equal initial mass.
In the simulations of \cite{Abate} the synthetic population 
of CEMP-$s$ stars has typically accreted a mass of approximately $0.1 M_{\odot}$.
Stars with 
initial mass $0.8 M_{\odot}$ at low metallicity have a lifetime of approximately
the age of the Universe;
however, if they experienced binary mass-transfer they would become high-carbon
band CEMP-$s$
stars of mass $\approx 0.9 M_{\odot}$ and consequently their lifetimes would
be shorter than 12 Gyr.
Hence, they would not be visible any longer. Similarly, stars of initial
mass of $0.7M_{\odot}$
would reach masses of around $0.8 M_{\odot}$ and still be visible after 12
Gyr, but for a shorter
time than single stars of the same initial mass. 
Also, the results of \cite{Moe} suggest that large initial mass ratios $q=M_{2,\mathrm{i}}/M_{1,\mathrm{i}}$
could be slightly more likely among low-mass binaries, hence increasing the
proportion of systems with initial
secondary mass $M_{2,\mathrm{i}}>0.8 M_{\odot}$.
The probability of observing a star at a given distance during its evolution
is 
proportional to the time its  apparent luminosity is above our detection
threshold.  If a star is visible in its main sequence phase, it will also
be visible in the RGB phase, but not once it reaches the white dwarf stage,
unless it is very near to us.
A faster evolving star  reaches  the white dwarf stage more quickly
and  is thus less likely to be observed. 
The decrease in lifetime of the high-carbon band population, due to their
increase in mass, naturally occurs at every metallicity, 
but its consequences become more evident at [Fe/H]$\leq-3.5$ because fewer
stars are known in this range.

 The determination
of the binary fraction and the period distribution for different metallicity
bins 
down to the lowest metallicities may provide a fundamental insight into the
properties of star formation at different metallicities. 

\subsection{Lithium abundances}

In Fig. \ref{plot_Li} we show the Li abundances as a function of [Fe/H]
for our programme stars and  for stars taken from the literature.
We classify the stars in three groups: low-carbon band CEMP stars (star symbols);
high-carbon band CEMP stars (crossed squares); and carbon normal stars (filled
hexagons), which  includes all stars with [C/Fe]$< 1.0$ or for which
the C abundance has not been measured. 
From this figure it appears that the Li abundances in the low-carbon band
CEMP stars have a behaviour
that is indistinguishable from that of the carbon normal stars.
On the other hand the high-carbon band CEMP stars are preferentially Li-depleted,
although there are a few measurements on the Spite plateau.
This is consistent with our hypothesis that the high-carbon 
CEMP stars are the result of mass transfer from an AGB companion. 
In most cases Li is depleted as a result of processing in the H-burning AGB
shell,
although in some cases it may also be produced through
the Cameron--Fowler mechanism \citep{CF71}.
Mass transfer from an AGB companion has been invoked 
to explain the occurrence of Li-rich main sequence stars
in globular clusters \citep{koch11,monaco12}, although pollution
from  nearby RGB stars having undergone the Li-flash is
an alternative explanation \citep{Pasquini14}.
The point is that transfer from an AGB always results
in an enhancement of carbon, but  Li may be either depleted
or enhanced.

The stars SDSS\,J1034+0701 and  SDSS\,J1247--0341 both have  
[Fe/H]\,$\sim -4.0 $. The measurements of the equivalent width
of the \ion{Li}{i} resonance doublet are very uncertain. In Fig. \ref{plot_lim4}
we show the observed spectra compared to a 3D NLTE synthetic spectrum. 
We stress that this is not a fit; we just plot the synthetic spectrum to
provide guidance to the position and expected strength of the 
 \ion{Li}{i} resonance doublet. Although
the measurement is uncertain, we believe that the detection of Li is robust.
Taken at face value, our measurements  place both stars
on the Spite plateau. All the other stars have either an upper limit
or a measurement that places them well below the Spite
plateau. While this behaviour, also referred to as `meltdown' of the Spite
plateau
(\citealt{sbordone10}, but see also \citealt{bonifacio07,aoki09}), is well
known,
these two stars on the Spite plateau appear exceptional.   
In fact, recently \citet{Matsuno17} pointed out that 
`no star in the literature
has comparable Li abundance to the Spite plateau below [Fe/H] of--3.5, except
for the primary of the 
double-lined binary system CS\,22876--032.'
The two components of CS\,22876--032 are shown as $\times$ symbols in
Figs.\,\ref{plot_Li} and \ref{plot_Li_teff}.
While SDSS\,J1034+0701 and  SDSS\,J1247--0341 have
an effective temperature well above 6000\,K, as does
CS\,22876-032\,A \citep{jonay08},  the star SDSS\,J0140+2344 has an effective
temperature below 6000\,K; 
it has roughly the same metallicity ([Fe/H]=--4.0) and,
like  CS\,22876-032\,B, it has a Li abundance well below the Spite plateau.
Therefore, our observations call into question the claim by
\citet{Matsuno17}, that Li abundances in the interval
$\rm -4.5 \le [Fe/H] \le -3.5$ are almost constant at a value
lower than the Spite plateau. 
For stars with \teff $> 6000$\,K,
there are some stars on the Spite plateau, 
at least down to [Fe/H]=--4.0. On the 
other hand, among the stars with \teff $\le $6000\,K and
[Fe/H]$\le -3.5$, lithium is always below
the Spite plateau. This can be seen in Fig.\,\ref{plot_Li_teff},
where the only star at effective temperature below 6000\,K
on the Spite plateau is SDSS\,J0907+0246 ([Fe/H]=--3.44).
The current observations  suggest that
the \teff\ for which lithium destruction in the stellar
atmospheres becomes important, increases as metallicity decreases,
although it is unlikely that this is the only effect that drives Li destruction
since there are several
stars hotter than 6000\,K  that are found well below the Spite plateau. 
\citet{BM97}  argued that for \teff $\ge $5700\,K Li destruction
was negligible; however, their  sample had only three stars
with metallicity below --3.0. From Fig.\,\ref{plot_Li_teff} it is clear 
that with increasing \teff\,Â there are fewer stars below the Spite plateau,
and, at any rate, the average Li abundance is higher. 

Our detection of Li in SDSS\,J1035+0641 is to date the measurement of Li
in the most
iron-poor unevolved star. It poses a strong constraint on any theory that
aims
to explain the cosmological Li problem. Any theory involving stellar destruction
of Li must accommodate the fact that in SDSS\,J1035+0641 Li is measurable.

In other stars of comparable iron abundance, it is either completely
destroyed or, at least, more severely depleted. 
Alternatively, if one tries to explain the Li abundance of this star
by a prompt enrichment in Li, one has to explain why such a prompt enrichment
has not taken place in the other stars with [Fe/H]$\le -5.0$.

\subsection{Abundances of other elements}

One of the main goals of the UVES observations was to increase the chemical
inventory 
of the stars at very low metallicity. 
The data presented in Table\,\ref{abbo2} fulfil this goal, 
although the low metallicities of the stars imply that we have many
upper limits. We have two measurements
of Co that confirm that Co is enhanced over Fe in EMP stars, as already shown
by \citet{cayrel04} and \citet{bonifacio09}.  In the two cases where Sr is
measured,
it is enhanced over Fe; however, this is certainly a selection effect since
for a given
effective temperature, the higher  the abundance, the stronger  the line
and thus the
easier it is to detect it.

In Fig.\,\ref{plot_alpha} we provide the ratios of the measured $\alpha$
elements, Mg, Si, and Ca, to Fe.  
In the upper panel, SDSS\,J1349+1407 clearly stands out as being extremely
enhanced in Mg, although its Ca abundance seems quite normal. There is another
star in our sample, SDSS\,J1050+2421 \citep{topos1},  that is highly enhanced
in Mg, 
although not as much as SDSS\,J1349+1407. 
We note that there is a discrepancy of about 1\,dex in the Ca abundance 
of SDSS\,J1050+2421 derived from the \ion{Ca}{i} resonance line
or from the \ion{Ca}{ii} lines. This discrepancy 
may be partly due to an incorrect gravity, since a 
discrepancy of about 0.5\, dex is found between the iron abundances derived
from the \ion{Fe}{i}
and \ion{Fe}{ii} lines \citep{topos1}. 
Based on
its proper motion, it has been pointed out that this star 
is most probably a main sequence star with  a \glog\ $\sim$ 4.7, rather
than 4.0,
as was assumed in \citet{topos1} (M. Bessell, private communication). 
The [\ion{Ca}{i}/\ion{Fe}{i}]
and the    [\ion{Ca}{ii}/\ion{Fe}{ii}] ratios differ by `only' about 0.6\,dex.

Thus, a change in gravity may alleviate the problem, but not solve it.
In Fig.\,\ref{plot_alpha} we plot the value
derived from the  \ion{Ca}{i} resonance line. Had we chosen the value derived
from \ion{Ca}{ii}, SDSS\,J1050+2421 would be highly enhanced in Ca as well.

The abundances of SDSS\,J1349+1407 resemble very closely those
of CS\,22949-037 \citep{depagne,norris02}; in fact, they have almost the
same [Mg/Ca]  ($\sim 1.2$).
Among the EMP stars, a [Mg/Ca]  almost as high is
found for HE\,1327-2326 (\citealt{frebel08} [Mg/Ca]=0.99) and an even higher
value is found
in SMSS\,J0313-6708  (\citealt{Keller14}) [Mg/Ca]=2.95).
Both CS\,22949-037 and HE\,1327-2326  are enhanced in Na: [Na/Fe]=+1.57 \citep{Andrievski}
and [Na/Fe]=+0.94 \citep{frebel08}.
Star  HE\,1012-1540,   analysed at high resolution by  the 0Z project 
\citep{Cohen13}, also seems
to belong to this group ([Mg/Ca]=+0.83, [Na/Fe]=+1.02).
Quite intriguingly, a star with similar properties has been found
in the Hercules ultra faint dwarf galaxy \citep{Koch08}.
For Her-2, with [Mg/Ca]=+0.93 and [Na/Fe]=+0.78, \citet{Koch08}
argued that this may be understood in terms of enrichment from
very massive stars with masses of the order of 30 M\sun.

The other feature that stands out in Fig.\ref{plot_alpha}
is that we find additional stars with low $\alpha$-to-iron ratios, 
for example SDSS\,J1442--0015 and SDSS\,J1247--0341.
This confirms our  earlier claim \citep{gto12}, based on X-Shooter
spectra, regarding the existence of a population of extremely metal-poor
$\alpha$-poor stars. 
The presence of metal-poor stars with low  $\alpha$-to-iron ratios,
in some cases quite extreme \citep{Ivans03}, is also supported 
by other investigations \citep{Cohen13,Susmitha16}. Recently
 \citet{hayes17} also recognised the presence of an $\alpha$-poor population
of stars belonging to the Galactic Halo,
albeit at a higher metallicity regime than the one considered in this paper.
The presence of these  $\alpha$-poor stars   has so-far
received little theoretical attention. 
While \citet{Ivans03} has invoked a prompt enrichment by 
type Ia supernovae, \citet{Cohen13} has suggested 
that the diversity in nucleosynthesis products of 
extremely metal-poor type II supernovae is greater than
that predicted by current models.
These stars could have formed in low-mass  dwarf galaxies and
subsequently accreted to the Milky Way Halo, carrying memory of 
a different chemical evolution, as has been claimed 
by \citet{hayes17}.

\section{Conclusions}

We have provided detailed abundances for a sample of seven EMP turn-off stars,
based on high-resolution spectra. 
All the stars for which carbon could be measured are CEMP stars 
and belong to the low-carbon band. 
For the most iron-poor star,
SDSS\,J1035+0641, we are unable to detect any iron lines, but we 
were able to place a stringent upper limit at [Fe/H]$\le -5.2$.
It is remarkable that we measure Li in this star, albeit
well below the Spite plateau. This is the first measurement of Li in 
an unevolved star with [Fe/H]$\le -5$, and it places a strong constraint on any theory
to explain the cosmological Li problem.
Three of the programme stars have [Fe/H]$\sim -4.0$, and  lithium
has been measured in all three. Quite interestingly, the Li abundance of
the two warmer stars 
places them squarely on the Spite plateau, while the cooler one lingers below.
This situation is reminiscent of what is observed in the binary system
CS\,22876-32 \citep{jonay08}. We have argued that this suggests that
the effective temperature at which Li depletion begins increases
with decreasing [Fe/H].

We confirmed that SDSS\,J1349+1407 is extremely enhanced in Mg but not
in Ca, and in fact its abundance pattern resembles that of CS\,22949-037
and
its Mg/Ca  is also similar to that of HE\,1327-2326.
This may suggest that there is a class of Mg-rich stars among EMP stars.

Our high-resolution observations have confirmed the existence of stars
with low $\alpha$-to-iron ratios, as pointed out by \citet{gto12}, based
on medium-resolution X-Shooter spectra. 
The TOPoS project has been very successful in increasing the number of  stars
with detailed abundances at and below a [Fe/H] of --4.0, yet the numbers
are still 
smaller than desirable for making robust statistical inferences. 
Hopefully the Pristine survey \citep{Pristine1,Pristine2,Pristine3}
and its massive follow-up with WEAVE \citep{WEAVE} will soon boost the numbers,
and the LAMOST survey as well \citep{Li2015a,Li2015}. 

\begin{table*}
\caption{Stellar parameters and abundances of C and Li}
\label{param}
\begin{center}
\begin{tabular}{lclrrcccll}
\hline \noalign{\smallskip}
 Star                   & \teff\ & \logg\ & $\xi$ & [Fe/H] & A(C) & A(C) & A(Ca) & A(Li) & EW(Li)\\
                        & K      & [C's]  & Km/s  &        &  1D  &  3D  & \ion{Ca}{ii}-K & & pm \\
\hline \noalign{\smallskip}
SDSS\,J0140+2344    & 5848   &  4.0   & 1.5   & \phantom{$<$}$-4.00\pm 0.14$ & 6.33 & 6.11 &      & \phantom{$<$}1.86 &\phantom{$<$}2.2 \\ 
SDSS\,J1034+0701    & 6224   &  4.0   & 1.5   & \phantom{$<$}$-4.01\pm 0.14$ & 6.28 & 5.96 &      & \phantom{$<$}2.24:&\phantom{$<$}2.8: \\ 
SDSS\,J1035+0641    & 6262   &  4.0   & 1.5   & $<-5.20$                     & 7.16 & 6.95 & 1.5  & \phantom{$<$}1.90  &\phantom{$<$}1.3\\ 
SDSS\,J1247--0341   & 6332   &  4.0   & 1.5   & \phantom{$<$}$-4.01\pm 0.32$ & 6.64 & 6.35 &      & \phantom{$<$}2.20:&\phantom{$<$}2.2:\\ 
SDSS\,J1349+1407    & 6112   &  4.0   & 1.5   & \phantom{$<$}$-3.60\pm 0.12$ & 7.00 & 6.82 &      & $< 0.9$ & $< 0.3$ \\
SDSS\,J1442--0015   & 5850   &  4.0   & 1.5   & \phantom{$<$}$-4.37\pm 0.16$ & 6.02 & 5.71 &      & $<1.58$ &$<1.2$   \\
SDSS\,J1507+0051    & 6555   &  4.0   & 1.5   & \phantom{$<$}$-3.41\pm 0.20$ &      &      & $<2.19$ &$<1.6$    \\  
\noalign{\smallskip}
\hline
\end{tabular}
\end{center}
\end{table*}

\begin{table*}
\caption{Abundances relative to hydrogen [X/H] with solar abundances from \citet{lodders09} 
for all elements except Fe and C whose solar abundances are from \citet{abbosun}.}
\label{abbo2}
\begin{center}
\begin{tabular}{llllllllll}
\hline
\noalign{\smallskip}
El          & Sun  &  J0140+2344     & J1034+0701     & J1035+0641                   & J1247--0341     & J1349+1407      & J1442--0015     & J1507+0051      \\
\hline \noalign{\smallskip}                                                                               
\ion{C}{i}  & 8.50 &  $-2.17$        & $-2.22$        &  \phantom{$<$}$-1.33$        & $-1.86$        & $-1.50$         & $-2.48$        &                 \\
\ion{Na}{i} & 6.30 &  $-4.22$        &                &                              &                & $-2.74\pm 0.12$ &                &                 \\
\ion{Mg}{i} & 7.54 &  $-3.37\pm 0.02$& $-3.28\pm 0.10$&                              & $-3.77\pm 0.23$& $-2.18\pm 0.11$ & $-4.19\pm 0.05$& $-3.30\pm 0.13$ \\
\ion{Al}{i} & 6.47 &  $-4.29$        & $-4.27$        &                              & $<-4.23$       & $-3.97$         & $<-4.68$       & $<-4.03$       \\
\ion{Si}{i} & 7.52 &  $-3.91$        & $-4.12$        &                              & $-4.27$        &                 & $-4.53$        & $-3.67$         \\
\ion{Ca}{i} & 6.33 &  $-2.79$        & $-4.17$        &                              &                & $-3.42\pm 0.16$ & $-4.48$       & $-3.35$         \\
\ion{Ca}{ii}& 6.33 &  $-3.66$        &                &  \phantom{$<$}$-4.83$        & $-4.07$        &                 &                & $-3.05$         \\
\ion{Sc}{ii}& 3.10 &                 & $-3.38$        &                              &                & $-3.54$         &                &                 \\
\ion{Ti}{i} & 4.90 &                 &                &                              &                &                 & $-2.33$        & $-3.04\pm 0.12$ \\
\ion{Ti}{ii}& 4.90 &  $-3.75\pm 0.14$& $-3.84\pm 0.04$&                              &                & $-3.22\pm 015$  & $-2.44$        &                 \\
\ion{Cr}{i} & 5.64 &  $-4.17\pm 0.12$&                &                              &                & $-3.73$         &                &                 \\
\ion{Mn}{i} & 5.64 &                 &                &                              &                & $-3.87\pm 0.03$ &                &                 \\
\ion{Fe}{i} & 7.52 &  $-4.00\pm 0.14$& $-4.01\pm 0.14$&  $<-5.20$                    & $-4.01\pm 0.32$& $-3.60\pm 0.12$ & $-4.37\pm 0.16$& $-3.41\pm 0.20$ \\
\ion{Fe}{ii}& 7.52 &                 &                &                              &                & $-3.58\pm 0.16$ &                & $-3.56$         \\
\ion{Co}{i} & 4.92 &  $-3.13\pm 0.34$&                &                              &                & $-2.87$         &                &                 \\
\ion{Ni}{i} & 6.23 &  $-3.86\pm 0.12$& $-3.99$        &                              &                & $-3.43$         &                &                 \\
\ion{Sr}{ii}& 2.92 &  $-3.43\pm 0.06$&                &                              &                &                 &                & $-3.16\pm 0.02$ \\
\noalign{\smallskip}
\hline
\end{tabular}
\end{center}
\end{table*}



\begin{acknowledgements}
We are grateful to our referee Mike Bessell for his very thorough and useful
 report of our paper.
The project is supported by fondation MERAC.
N.C., H.G.L., and A.J.G. were supported by Sonderforschungsbereich SFB 881
``The Milky Way System'' (subproject A4) of the German Research
Foundation (DFG).
R.S.K. acknowledges support from the European Research Council via the Advanced
Grant 
``STARLIGHT: Formation of the First Stars'' (project number 339177). 
C.A. acknowledges funding from the Alexander von Humboldt Foundation
\end{acknowledgements}

\bibliographystyle{aa}

\end{document}